\documentclass[a4paper,pdftex,reqno,12pt]{amsart}

\allowdisplaybreaks[1]

\usepackage{geometry}

\usepackage{graphicx}
\usepackage{enumerate}
\usepackage{courier}
\usepackage{times}
\usepackage{amsmath,amssymb}

\usepackage{hyperref}

\theoremstyle{plain}
\newtheorem{Theorem}{Theorem}[section]
\newtheorem{Proposition}[Theorem]{Proposition}

\newtheorem{Corollary}[Theorem]{Corollary}
\newtheorem{Lemma}[Theorem]{Lemma}

\newenvironment{Proof}
{\begin{trivlist}\item[]{{\sc Proof.}}}{\hfill{$\square$}\noindent\end{trivlist}}

\theoremstyle{definition}
\newtheorem{Definition}[Theorem]{Definition}

\theoremstyle{remark}

\begin{document}


\title{Approximating power by weights}

\date{}

\author{Sascha Kurz}
\address{Sascha Kurz, University of Bayreuth, 95440 Bayreuth, Germany}
\email{sascha.kurz@uni-bayreuth.de}

\begin{abstract}
  Determining the power distribution of the members of a shareholder meeting or a legislative committee 
  is a well-known problem for many applications. In some cases it turns out that power is nearly proportional  
  to relative voting weights, which is very beneficial for both theoretical considerations and practical 
  computations with many members. We present quantitative approximation results with precise error bounds 
  for several power indices as well as impossibility results for such approximations between power and weights.
\end{abstract}

\maketitle

\section{Introduction}

Consider a stock corporation whose shares are hold by three major stockholders owning 
35\%, 34\%, and 17\%, respectively. The remaining 14\% are widely spread. 
Assuming that decisions are made by a simple majority rule, all three major stockholders have  
equal influence on the company's decisions, while the private shareholders have no say. To 
be more precise, any two major stockholders can adopt a proposal, while the private   
shareholders together with an arbitrary major stockholder need further affirmation. 
Such decision environments can be captured by means of weighted voting games. Formally, a 
weighted (voting) game consists of a set of players $N=\{1,\dots,n\}$, a vector of non-negative weights 
$w=(w_1,\dots,w_n)$, and a positive quota $q$. A proposal is accepted if and only if the weight sum 
of its supporters meets or exceeds 
the quota. 
Committees that decide between 
two alternatives have received wide attention. Von Neumann and Morgenstern introduced the notion of simple 
games, which is a super class of weighted games, in \cite{von1953theory}. Examples of decision-making bodies that 
can be modeled as weighted games are the US Electoral College, the Council of the European Union, the 
UN Security Council, the International Monetary Fund or the Governing Council of the European Central Bank.  
Many applications seek to evaluate players' influence or power in simple or weighted games, see, e.g., 
\cite{leech1987ownership}. 
The initial example illustrates that shares or 
weights can be a poor proxy for the distribution of power. Using the taxicab metric, i.e., the 
$\Vert\cdot\Vert_1$-distance, the corresponding distance between shares and relative power is 
$\left|0.35-\tfrac{1}{3}\right|+\left|0.34-\tfrac{1}{3}\right|+\left|0.17-\tfrac{1}{3}\right|+
\left|0.14-0\right|\approx$32.67\%.  

If the weights add up to one, then we speak of relative weights.
The insight that the power distribution differs from 
relative weights, triggered the invention of so-called power indices like the Shapley-Shubik index \cite{shapley1954method},  
the Penrose-Banzhaf index \cite{banzhaf1964weighted}, or the nucleolus \cite{schmeidler1969nucleolus}. Due to 
the combinatorial nature of most of those 
indices, qualitative assessments are technically demanding and large numbers of involved parties cause computational 
challenges \cite{chalkiadakis2011computational}. Moreover, there is a large variety of different power indices proposed so far. 
On the positive side, there are a few \textit{limit results}, which state that, under 
certain technical conditions, the power distribution of an infinite sequence of weighted voting 
systems tends to the relative weight distribution. This of course simplifies the analysis. The aim 
of this paper is to provide quantitative results of the form 
\begin{equation}
  \Vert p-w\Vert_1=\sum_{i=1}^n \left| p_i-w_i\right| \le \frac{c\Delta}{\min\{q,1-q\}},
  \label{ie_general_form} 
\end{equation}
where $w$ is the relative weight vector, $q$ the quota, $p$ the power distribution 
induced by a certain power index, $\Delta=\max_i w_i$ the maximum relative weight, and $c\in\mathbb{R}_{>0}$ 
a constant depending on the chosen power index. This 
inequality 
provides a concrete error bound based on just a few invariants of the underlying weighted game. Although limit results for 
sequences of weighted games can be derived in general, Inequality~(\ref{ie_general_form}) can also be applied to 
a single weighted game. 
Applications range from approximating power distributions 
with many involved parties, where the exact evaluation is computationally infeasible, to statements 
about power distributions in situations with incomplete or uncertain information. In our above example 
there may be 
many private shareholders whose exact shares are either unknown (due to the lack of a 
reporting obligation) or highly volatile. (Our precise statement about the exact power distribution, independent 
of the distribution of the widely spread shareholdings and almost independent of the chosen power index, 
is due to a rather special situation, see the end of Subsection~\ref{subsec_games}.) Results for different 
distance measures can be derived in a unified way, which makes the choice of the $\Vert\cdot\Vert_1$-distance 
less special. The precise involvement of the invariants $\Delta$, $q$, and $1-q$ in the right hand side of 
Inequality~(\ref{ie_general_form}) is rather an explanation of a universal behavior than a limitation. 
We will derive lower bounds for the constant $c$ independent of the properties of the chosen power index, i.e., besides 
the constant, results like Inequality~(\ref{ie_general_form}) are the best we can hope for. This explains the 
necessity of several 
conditions used in known limit results.   

For the Shapley-Shubik index Neyman obtained in 1982:
\def\ordering{\varrho}
\begin{Theorem}\cite{neyman1982renewal}
  Let $n\in\mathbb{N}$, $N=\{1,\dots,n\}$, $0<q<1$, $w\in\mathbb{R}_{\ge 0}^n$ with $\Vert w\Vert_1=1$, and $P(i,q)$ 
  be the probability that in a random order of $N$, $i$ is the first element in the order for which the $w$-accumulated 
  sum exceeds $q$. For every $\varepsilon>0$ there exist constants $\delta>0$ and $K>0$ such that if 
  $\rho=\max_{i\in N} w_i<\delta$, and $K\rho<q<1-K\rho$ then $\Vert P(\cdot,q)-w\Vert_1<\varepsilon$. 
\end{Theorem}

In other words, if the maximum relative weight $\Delta$ approaches zero and the relative quota $q$ is not 
too near to the extreme points $0$ or $1$, then the power distribution tends to the vector of relative 
weights if the distance is measured by the $\Vert\cdot\Vert_1$-norm. The precise dependence of $\delta$ and 
$K$ on $\varepsilon$ is hidden in the technical lemmas of \cite{neyman1982renewal}. This is due to the fact 
that the original motivation behind this result was the study of the asymptotic value of non-atomic games. 

Another well-known limit theorem is the Penrose limit theorem (PLT). It is an unproven statement implicitly contained in 
\cite{penrose1952objective}. Loosely speaking it states that, for a quota of one half, under certain conditions, 
the ratio between the Penrose-Banzhaf indices of any two voters converges to the ratio between their weights as the 
number of voters increases and the maximum relative weight decreases to zero. In \cite{lindner2004ls} the authors used 
Theorem~1.1 in order to derive a version of the PLT for the Shapley-Shubik index for so-called replicative $q$-chains, 
where finitely many types of voters get replicated with a strictly positive frequency. In Lemma~\ref{lemma_quotient} 
we deduce a general PLT-type result from Inequality~(\ref{ie_general_form}),  
cf.\ \cite[Proposition 1]{kurz2014nucleolus}. Using a local central limit theorem (for normalized sums of 
independent random variables) Lindner and Machover, see \cite{lindner2004ls}, also obtained a PLT for the 
Penrose-Banzhaf index for $q=\frac{1}{2}\sum_{i=1}^n w_i$ and divisibility conditions on the involved 
(non-normalized) integer weights.  

Besides the Shapley-Shubik and the Penrose-Banzhaf index, further limit results have only been obtained for the 
nucleolus. In \cite[Lemma 1]{kurz2014nucleolus} the authors have proven Inequality~(\ref{ie_general_form}) 
for $c=2$, which implies a PLT-type result. 

\smallskip

One important problem in the context of power indices and weighted games, 
is the so-called {\lq\lq}inverse power index problem{\rq\rq}, see 
e.g.~\cite{de2017inverse,koriyama2013optimal,kurz2017democratic,kurz2014heuristic} and 
the references therein. It asks for weights and a quota such that the corresponding power distribution 
meets a given ideal power distribution as closely as possible. Since there is only a finite number of different 
weighted or simple games, 
it is obvious that some power vectors can not be approximated 
too closely if the number of voters is small. In \cite{alon2010inverse} Alon and Edelman showed 
that there are also vectors that are hard to approximate by the Penrose-Banzhaf index of a simple game 
if most of the mass of the vector is concentrated on a small number of coordinates. This goes in line 
with a relatively large maximum weight $\Delta$. Generalizations and impossibility results for other 
power indices have been obtained in \cite{kurz2016inverse}.

The case where the power distribution coincidentally matches relative weights has received special attention in the 
literature. For the Penrose-Banzhaf index, the subclass of spherically separable simple games has this property, 
see \cite{houy2014geometry}. In \cite{peleg1968weights} Peleg shows that a homogeneous constant-sum weighted game 
has a nucleolus which equals the relative homogeneous weights. A similar result for the nucleolus of weighted games 
with many replicated voters that have integral weights from a finite set has been obtained in 
\cite[Proposition 2]{kurz2014nucleolus}. 

For more practically orientated studies on the ownership and control structure of stock companies we refer the 
interested reader, e.g., to \cite{leech1991ownership}. Algorithms to approximate power indices can, e.g., be found in 
\cite{bachrach2010approximating,fatima2008linear,leech2003computing}. The expected difference 
between power and weights 
has been studied in \cite{jelnov2014voting} for the Shapley-Shubik and the Penrose-Banzhaf index. 
Intervals for the power of voters in weighted games with uncertain weights can also be computed with integer linear 
programming techniques, see \cite{kurz2012inverse}. However, these techniques (currently) are computationally 
infeasible for more than, 
say, 20~voters.

\smallskip

The rest of the paper is organized as follows. In Section~\ref{sec_preliminaries} we introduce 
weighted games, power indices, norms and limits. 
Our main results are concluded in Section~\ref{sec_main_results}, which is subdivided into three subsections. First we discuss 
invariants of weighted games that are suitable for upper bounds on the deviation of power and relative weights. 
Proposition~\ref{prop_impossible} shows that the relative quota, the number of voters, or the maximum relative weight alone 
are impractical for this purpose. In Theorem~\ref{thm_lower_bound} we argue that bounds of the form of 
Inequality~(\ref{ie_general_form}) are the best that we can hope for (in a certain restricted sense). One such upper bound, which is applicable 
for the nucleolus, is obtained in Lemma~\ref{lemma_qdelta_bound_winning}. A more general result applicable to a larger class of 
power indices is presented in Theorem~\ref{thm_main}. We close this section by numerical investigations 
for other power indices from the literature that are not captured by the mentioned theorem. In Section~\ref{sec_implications} we 
briefly show how quotient-like statements as 
PLT-type results can be obtained in general from those $\Vert\cdot\Vert_1$ upper bounds. We close with a 
conclusion and future research directions in Section~\ref{sec_conclusion}. The more technical proofs from the results 
of Section~\ref{sec_main_results} are 	moved to Section~\ref{sec_evacuated_proofs} in the appendix.    

\section{Preliminaries}
\label{sec_preliminaries}

This section collects some notation and basic facts. Simple games, weighted games and power indices as a tool for 
the measurement of voting power are treated in Subsection~\ref{subsec_games}. Subsection~\ref{subsec_limits} treats the 
mathematical basics of distance functions, norms, and limits. 

\subsection{Weighted games, simple games and measurement of power}
\label{subsec_games}

For a positive integer $n$ let $N=\{1,\dots, n\}$ be the set of voters. A \emph{simple game} is a 
mapping $v\colon 2^N\to\{0,1\}$ from the subsets of $N$ to binary outcomes satisfying 
$v(\emptyset)=0$, $v(N)=1$, and $v(S)\le v(T)$ for all $\emptyset\subseteq S\subseteq T\subseteq N$. 
The interpretation in the context of binary voting systems is as follows. A subset $S\subseteq N$ 
is considered as the set of voters that are in favor of a proposal, i.e., which vote {\lq\lq}yes{\rq\rq}.  
$v(S)$ encodes the group decision, i.e., $v(S)=1$ if the proposal is accepted and $v(S)=0$ otherwise.  
So, these assumptions 
are quite natural for a voting system with binary options in the 
input and output domain. A simple game $v$ is called \emph{weighted} if there exist weights $w\in\mathbb{R}_{\ge 0}^n$ 
and a quota $q\in\mathbb{R}_{>0}$ such that $v(S)=1$ if and only if $w(S):=\sum_{i\in S} w_i\ge q$.\footnote{Some 
authors require $w(S)>q$, which may be written as $w(S)\ge q'$ for $q'$ slightly larger than $q$.} From the conditions 
of a simple game we conclude $0<q\le w(N)$. If $w(N)=1$ we speak of normalized or relative weights, where 
$0<q\le 1$. We denote the respective game by $v=[q;w]$ and refer to the pair $(q;w)$ as a weighted 
representation, i.e., we can have $[q;w]=[q';w']$ but $(q;w)\neq (q';w')$. The example from the introduction 
can, e.g., be represented by $(51\%;35\%,34\%,17\%,14\%)$, $\left(\tfrac{1}{2};\tfrac{1}{3},\tfrac{1}{3},\tfrac{1}{3},0\right)$, 
or $(6;4,3,3,1)$, where the fourth voter mimics the private shareholders.
Two voters $i,j\in N$ are called \emph{equivalent} if $v(S\cup\{i\})=v(S\cup\{j\})$ for all 
$S\subseteq N\backslash\{i,j\}$. If $v(\{i\})=1$, we call voter $i$ a \emph{passer} and a \emph{null voter} 
if $v(S\cup\{i\})=v(S)$ for all $S\subseteq N\backslash\{i\}$. 

A \emph{power index} $\varphi$ is a family of mappings from the set of simple or weighted games on $n$ voters 
into $\mathbb{R}^n$, where $\varphi_i(v)$ denotes the $i$th component of $\varphi(v)\in\mathbb{R}^n$. We call $\varphi$ 
\emph{positive} if $\varphi(v)\in\mathbb{R}_{\ge 0}^n\backslash\{0\}$ for all possible games $v$. We say that 
$\varphi$ satisfies the \emph{null voter property} if $\varphi_i(v)=0$ for each null voter $i\in N$. We call $\varphi$ \emph{symmetric} 
if $\varphi_i(v)=\varphi_j(v)$ for all equivalent voters $i,j$ and \emph{efficient} if $\Vert\varphi(v)\Vert_1=1$ 
for all possible games $v$. However, if $\varphi$ is not efficient but positive, then 
$\varphi_i'(v):=\varphi_i(v)/\sum_{j=1}^n \varphi_j(v)$ is both. The absolute Penrose-Banzhaf index is defined 
by $\frac{1}{2^{n-1}}\sum_{S\subseteq N\backslash \{i\}} \left(v(S\cup\{i\})-v(S)\right)$ for voter $i\in N$. With this, 
the (relative) Penrose-Banzhaf index is the corresponding efficient version as constructed before. The 
Shapley-Shubik index for voter~$i$ is given by 
$\sum_{S\subseteq N\backslash\{i\}} \frac{|S|!\cdot(n-|S|-1)!}{n!}\cdot\left(v(S\cup\{i\})-v(S) \right)$. In order 
to define the \emph{nucleolus} of a simple game we need some preparations. In our context, an imputation $x$ 
is an element of $\mathbb{R}_{\ge 0}^n$ with $\Vert x\Vert_1\le 1$. For an imputation $x$ and $S\subseteq N$ 
we call $e(S,x)=v(S)-x(S)$ the \emph{excess}, where $x(S)=\sum_{i\in S}x_i$. With this, the \emph{excess vector} 
is the weakly monotonically decreasing list of the excesses of the $2^n$ subsets of $N$. E.g., for $v=[4;3,2,1,1]$ and 
$x=\left(\tfrac{1}{3},\tfrac{1}{3},\tfrac{1}{6},\tfrac{1}{3}\right)$ the excess vector is given by
$$
 \left(\tfrac{1}{2},\tfrac{1}{2},
 \tfrac{1}{3},\tfrac{1}{3},\tfrac{1}{3},
 \tfrac{1}{6},\tfrac{1}{6},
 0,0,
 -\tfrac{1}{6},-\tfrac{1}{6},
 -\tfrac{1}{3},-\tfrac{1}{3},-\tfrac{1}{3},
 -\tfrac{1}{2},-\tfrac{1}{2}\right).
$$    
The (unique) imputation $x^\star$ that yields the lexicographical minimal excess vector is called the \emph{nucleolus} 
of $v$. See \cite{schmeidler1969nucleolus} for the original definition which does not apply to simple games with 
more than one passer. $\Vert x^\star\Vert_1=1$ is automatically satisfied by the excess minimizer. Note that some authors require 
$\Vert x \Vert_1=1$ for any imputation. We remark that all three mentioned power indices are positive, symmetric, efficient and 
satisfy the null voter property. 

In order to describe the structure underlying the example from the introduction, we have to introduce 
an unanimity game $u_S$ as follows: $u_S(T)=1$ if and only if $S\subseteq T$, where $\emptyset \neq S\subseteq N$. For each 
symmetric and efficient power index $\varphi$ satisfying the null voter property we have $\varphi_i(u_S)=1/|S|$ 
for all $i\in S$ and $\varphi_i(u_S)=0$ otherwise. 

\subsection{Mathematical basics of limits, norms and distance functions}
\label{subsec_limits}
A \textit{distance function} or \textit{metric} is used to measure the \textit{distance} between two elements $x,y$ of some (arbitrary) set $U$.
For a metric we assume no structure of the set $U$, which in turn allows a vast diversity of different 
metrics in general. Given a metric $d$ on a set $U$ we can compare any two elements of $U$ according to their 
distance. For a \emph{sequence} $\left(x_n\right)_{n\in\mathbb{N}}$, i.e., an infinite ordered list of elements 
$x_n\in U$, 
in $U$, we can formalize the idea of the $x_n$ \emph{tending} to some ultimate
$x\in U$ as follows:
\begin{Definition}
  Given a metric space $(U,d)$, i.e., a set $U$ and a metric $d$ on $U$, we say that $x\in U$ is the \emph{limit} of a 
  sequence $\left(x_n\right)_{n\in\mathbb{N}}$ (in $U$) if for all $\varepsilon\in\mathbb{R}_{>0}$ there exists an $N(\varepsilon)\ge 0$ 
  such that for all integers $n\ge N(\varepsilon)$, we have $d(x_n,x)<\varepsilon$. If a sequence admits a limit, we say that the 
  sequence is \textit{convergent}. 
\end{Definition}  
We remark that each convergent sequence uniquely determines a limit. However, whether a sequence converges  
can depend on the used metric. 
We thus restrict attention to metrics induced by a norm of a finite dimensional vector space.
Each norm $\Vert\cdot\Vert$ induces a distance function via $d(x,y):=\Vert x-y\Vert$. 
For $V=\mathbb{R}^n$ examples of norms are given by $\Vert x\Vert_1=\sum_{i=1}^n \left|x_i\right|$ and $\Vert x\Vert_\infty 
=\max_{1\le i \le n} \left| x_i\right|$. Given a vector space $V$ two metrics $\Vert\cdot\Vert$ and $\Vert\cdot\Vert'$ are 
called \emph{equivalent} if there exist $l_1,l_2,u_1,u_2\in\mathbb{R}_{>0}$ such that $l_1\Vert v\Vert \le \Vert v\Vert'\le u_1 \Vert v\Vert$ 
and $l_2\Vert v\Vert' \le \Vert v\Vert\le u_2 \Vert v\Vert'$ for all $v\in V$. In a finite-dimensional vector space 
all norms are equivalent. As an example consider
$$
  1\cdot \Vert x\Vert_\infty\le \Vert x\Vert_1\le n\cdot\Vert x\Vert_\infty
  \quad\text{and}\quad
  \frac{1}{n}\cdot \Vert x\Vert_1\le \Vert x\Vert_\infty\le 1\cdot\Vert x\Vert_1 
$$
for all $n\in\mathbb{N}_{>0}$ and all $x\in\mathbb{R}^n=:V$. (Indeed the stated constants are tight as they are attained at $x=(1,0,\dots,0)$ 
and $x=(1,\dots,1)$.) So, in $\mathbb{R}^n$ a sequence is convergent with respect to a metric induced by norm $\Vert\cdot\Vert$ if and only 
if it is convergent with respect to a metric induced by another norm $\Vert\cdot\Vert'$, i.e., there is no need to explicitly state the 
used norm. (As long as the application does not call for a specific metric that is not induced by a norm or the dimension $n$ of 
the ambient space is varying too.)  

The bound 
$\Vert x\Vert_{\infty}\le\Vert x\Vert_1$ can be slightly improved in our context.

\begin{Lemma}
  \label{lemma_improved_relation_infty_1}
  For $w,\overline{w}\in\mathbb{R}^n_{\ge 0}$ with $\Vert w\Vert_1=\Vert \overline{w}\Vert_1=1$, we have 
  $\Vert w-\overline{w}\Vert_{\infty}\le \frac{1}{2}\Vert w-\overline{w}\Vert_1$.
\end{Lemma}
\begin{Proof}
  With $S:=\{1\le i\le n\mid w_i\le \overline{w}_i\}$ and $A:=\sum_{i\in S} \left(\overline{w}_i-w_i\right)$, 
  $B:=\sum_{i\in N\backslash S} \left(w_i-\overline{w}_i\right)$, where $N=\{1,\dots,n\}$, we have $A-B=0$ since $\Vert w\Vert_1=\Vert \overline{w}\Vert_1$ 
  and $w,\overline{w}\in\mathbb{R}_{\ge 0}^n$. Thus, $\Vert w-\overline{w}\Vert_1=2A$ and $\Vert w-\overline{w}\Vert_{\infty}\le \max\{A,B\}=A$.
\end{Proof}

\section{Inequalities between weights and power indices}
\label{sec_main_results}
We are interested in upper bounds for the distance between the relative weights $w$ of a weighted game $[q;w]$ (with $n$ voters) and the 
corresponding power distribution $\varphi([q;w])$. As argued in the previous subsection, we should limit our considerations on distance functions 
induced by a norm $\Vert\cdot\Vert$. While any two norms are equivalent for a fixed dimension $n$, the corresponding constants can of course 
depend on $n$. So, we have to explicitly state which norms we want to use. Here, we restrict ourselves onto the norms 
$\Vert\cdot\Vert_1$ and $\Vert\cdot\Vert_\infty$, which 
have nice mathematical and algorithmic properties. Note that $\Vert x\Vert_\infty\le\Vert x\Vert_2\le \Vert x\Vert_1$, so that 
we capture two kinds of \textit{extreme} positions. Knowing 
$w$, $[q;w]$, $\varphi(\cdot)$ and $\Vert\cdot\Vert$ of course uniquely determines $\Vert w-\varphi\left(\left[q;w\right]\right) \Vert$.  
We thus aim at  deriving upper bounds only invoking few invariants of a given representation $(q;w)$ and the corresponding weighted game $[q;w]$. 
In Subsection~\ref{subsec_3_1} we briefly describe the invariants considered in this paper and discuss possible alternatives. The aim 
of Subsection~\ref{subsec_3_2} is to derive lower bounds for the distance between relative weights and power in the worst case. 
Upper bounds are treated in Subsection~\ref{subsec_3_3}.

\subsection{Invariants of weighted games and their representations}
\label{subsec_3_1}
We consider a weight\-ed game with normalized representation $(q;w)$, i.e., $w\in\mathbb{R}_{\ge 0}^n$ with $\Vert w\Vert_1=1$. Useful and 
easy invariants are the number of voters $n$, the quota $q\in(0,1]$, and the maximum weight $\max_i w_i=\Vert w\Vert_{\infty}$.\footnote{
For an arbitrary representation $(q;w)$ we consider the normalized quota $q/\sum_{i=1}^n w_i$ and the normalized relative weight 
$\max\{w_i/\sum_{j=1}^n w_j\mid 1\le i\le n\}$.} However, also more sophisticated invariants of weight vectors have been 
studied in applications. The so-called \textit{Laakso-Taagepera index} a.k.a.\ \textit{Herfindahl-Hirschman index}, 
c.f.\ \cite{laakso1981proportional}, is used in Industrial Organization to measure the concentration of firms in a market, see, e.g., 
\cite{curry1983industrial}.   
\begin{Definition}
  For $w\in\mathbb{R}_{\ge 0}^n$ with $w\neq 0$ the \textit{Laakso-Taagepera} index is given by
  $$
    L(w)
    =\left(\sum\limits_{i=1}^{n}w_i\right)^2 / \sum\limits_{i=1}^{n} w_i^2.
  $$
\end{Definition}

In general we have $1\le L(w)\le n$. If the weight vector $w$ is normalized, then the formula simplifies to 
$L(w)=1/\sum_{i=1}^n w_i^2$. Under the name {\lq\lq}effective number of parties{\rq\rq} the index is widely used in political science 
to measure party fragmentation, see, e.g., \cite{laakso1979effective}.  
We observe the following relations between the maximum relative weight 
$\Delta=\Delta(w)$ and the Laakso-Taagepera index $L(w)$:
\begin{Lemma}
  \label{lemma_relation_maximum_laakso_taagepera}
  For $w\in\mathbb{R}_{\ge 0}^n$ with $\Vert w\Vert_1=1$, we have
  $$
    \frac{1}{\Delta}\le
    \frac{1}{\Delta\left(1-\alpha(1-\alpha)\Delta\right)}\le L(w)\le
    \frac{1}{\Delta^2+\frac{(1-\Delta)^2}{n-1}}\le\frac{1}{\Delta^2}
  $$
  for $n\ge 2$, where $\alpha:=\frac{1}{\Delta}-\left\lfloor\frac{1}{\Delta}\right\rfloor\in[0,1)$. 
  If $n=1$, then  $\Delta=L(w)=1$. 
\end{Lemma}
\begin{Proof}
Optimize $\sum\limits_{i=1}^n w_i^2$ with respect to the constraints $w\in\mathbb{R}^n$, $\Vert w\Vert_1=1$, and $\Delta(w)=\Delta$, 
see Section~\ref{sec_evacuated_proofs} for the technical details.
\end{Proof} 

So, any lower or upper bound involving $L(w)$ can be replaced by a bound involving $\Delta$ instead. Since $\Delta$ has nicer analytical 
properties and requires less information on $w$, we stick to $\Delta$ in the following. We remark that there are similar inequalities for 
other indices measuring market concentration. 

In the context of the study of limit theorems for power distributions of weighted games the  
Laakso-Taagepera index was suggested in \cite{leech2013power}. However, the limit behavior of $L(w)$ is in 
one-to-one correspondence to the limit behavior of $1/\Delta(w)$: 

\begin{Corollary}
  \label{cor_relation_maximum_laakso_taagepera}
  Let $\left(w^m\right)_{m\in\mathbb{N}}$ be a sequence of vectors with non-negative entries and $\left\Vert w^m\right\Vert_1=1$. 
  (To be more precise, $w^m\in \mathbb{R}^{n_m}_{\ge 0}$ for some $n_m\in\mathbb{N}_{>0}$.) Then, we have 
  $$
    \lim\limits_{m\to\infty} \Delta(w^m)=0
    \quad\Longleftrightarrow\quad
    \lim\limits_{m\to\infty} L(w^m)=\infty.
  $$
\end{Corollary}

We leave the study of other possible invariants of weighted games and their representations for future research. 
Ideas for other invariants may, e.g., be smallest weight of non-null voters or moments from statistics.

\subsection{Lower bounds for the worst case approximation}
\label{subsec_3_2}

In order to study the question which set of invariants permits a meaningful upper bound on the distance between 
relative weights and power, we consider constructions meeting the prescribed invariants to obtain lower bounds 
on the worst case approximation.

Since a large number of power indices is introduced in the literature and this stream does not seem to dry out, it would be very desirable 
to have approximation statements which hold for large classes of power indices. With no assumption other than power being a function 
of the weighted game itself rather than its representation we observe: 
\begin{Lemma}
  \label{lemma_general_lower_distance_bound}
  Let $n\in\mathbb{N}_{>0}$, $q,\overline{q}\in(0,1]$, $w,\overline{w}\in\mathbb{R}^n_{\ge 0}$ with $\left\Vert w\right\Vert_1=
  \left\Vert \overline{w}\right\Vert_1=1$ and $[q;w]=[\overline{q};\overline{w}]$, $\Vert\cdot\Vert$ be an arbitrary norm on 
  $\mathbb{R}^n$ and $\varphi$ be a power index,  
  then we have 
  $$
    \max\left\{\left\Vert w-\varphi\left(\left[q;w\right]\right)\right\Vert,
    \left\Vert \overline{w}-\varphi\left(\left[\overline{q};\overline{w}\right]\right)\right\Vert
    \right\}\ge \frac{\left\Vert w-\overline{w}\right\Vert}{2}.
  $$
\end{Lemma} 
\begin{Proof}
  Using the triangle inequality yields $\left\Vert w-\varphi\left(\left[q;w\right]\right)\right\Vert+
    \left\Vert \overline{w}-\varphi\left(\left[\overline{q};\overline{w}\right]\right)\right\Vert\ge 
    \left\Vert w-\overline{w}\right\Vert$ from which we can conclude the stated 
    inequality. 
\end{Proof}

So, instead of lower bounds for the distance between relative weights and power, we will consider lower bounds for the maximum 
distance between two relative weight vectors of the same weighted game being compatible with the considered invariants. 
The general lower bound of Lemma~\ref{lemma_general_lower_distance_bound} will now be used to show that controlling the 
quota and the number of voters cannot yield reasonable limit results, i.e., there exist examples such that the distance between 
power and relative weights is lower bounded by a positive constant not depending on the number of voters.
\begin{Lemma}
  \label{lemma_lb_approximation_q}
  For each $q\in(0,1]$ there exists a weighted game $v=[q;w]=[q;\overline{w}]$ with $n\ge 2$ voters, where 
  $w,\overline{w}\in\mathbb{R}^n_{\ge 0}$, and $\Vert w\Vert_1=\Vert \overline{w}\Vert_1=1$, such that   
  $\Vert w-\overline{w}\Vert_{\infty}\ge \frac{1}{3}$ and $\Vert w-\overline{w}\Vert_{1}\ge \frac{2}{3}$. 
\end{Lemma} 
\begin{Proof}
  We give general constructions for different ranges of $q$:
  \begin{itemize}
    \item $\frac{2}{3}<q\le 1$: $w=\left(\frac{2}{3},\frac{1}{3},0,\dots,0\right)$, $\overline{w}=\left(\frac{1}{3},\frac{2}{3},0,\dots,0\right)$; 
    \item $\frac{1}{3}<q\le \frac{2}{3}$: $w=\left(\frac{2}{3},\frac{1}{3},0,\dots,0\right)$, $\overline{w}=\left(1,0,\dots,0\right)$;
    \item $0<q\le \frac{1}{3}$: $w=\left(\frac{2}{3},\frac{1}{3},0,\dots,0\right)$, $\overline{w}=\left(\frac{1}{3},\frac{2}{3},0,\dots,0\right)$.
  \end{itemize}
\end{Proof}

So, if we just know that the number of voters tends to infinity and the quotas are fixed to some arbitrary number in $(0,1]$ or some 
arbitrary subinterval of $(0,1]$, then no general limit result is possible. For a single weighted game we can state a constant number as 
a lower bound for the distance between relative weights and power independent of the invariants $q$ and $n$, both in the distances induced 
by the $\Vert\cdot\Vert_1$- and the $\Vert\cdot\Vert_\infty$-norm, respectively.  


Similarly, it is not sufficient to require that the maximum relative weight $\Delta$ tends to zero, which is equivalent to $L(w)\to\infty$ and 
implies that the number of voters grows without bounds. In terms of a single weighted game, we construct a weighted representation consisting 
of any number of voters that is sufficiently large and exactly meets the chosen value of $\Delta$. Then we construct another weighted representation 
of the same weighted game whose distance to the first weight vector is lower bounded by a constant in the distance induced by the 
$\Vert\cdot\Vert_1$-norm.  

\begin{Lemma}
  \label{lemma_lb_approximation_delta_1}
  For each $\Delta\in(0,1)$ there exists a weighted game $v=[q;w]=[q;\overline{w}]$ with $n\ge \frac{4}{3\Delta}+6$ voters, where 
  $q\in(0,1)$, $w,\overline{w}\in\mathbb{R}^n_{\ge 0}$, $\Delta(w)=\Delta(\overline{w})=\Delta$, and  
  $\Vert w\Vert_1=\Vert \overline{w}\Vert_1=1$, such that $\Vert w-\overline{w}\Vert_{1}\ge \frac{2}{3}$ and 
  $\Vert w-\overline{w}\Vert_{\infty}\ge\Delta/2$.
\end{Lemma}
\begin{Proof} 
  If $\Delta\ge \frac{2}{3}$, we can consider a weighted game with two passers and $n-2$ null voters. One representation is given by 
  $q=1-\Delta$ and $w=(\Delta,1-\Delta,0,\dots,0)$. Of course we can swap the weights of the first two voters and obtain a 
  second representation given by quota $q$ an weight vector $\overline{w}=(1-\Delta,\Delta,0,\dots,0)$. With this, we compute 
  $\Vert w-\overline{w}\Vert_{1}=2\cdot (2\Delta-1)\ge \frac{2}{3}$ and
  $\Vert w-\overline{w}\Vert_{\infty}=2\Delta-1\ge \Delta/2$. 

  If $0<\Delta<\frac{2}{3}$, we define an integer $a:=\left\lfloor\frac{2}{3\Delta}\right\rfloor\ge 1$ and consider a weighted game 
  with $2a$ passers and $n-2a$ null voters. One representation is given by $q=\Delta/2$, $w_{2i-1}=\Delta$, $w_{2i}=\Delta/2$ for $1\le i\le a$, 
  $w_{2a+1}=w_{2a+3}=w_{2a+5}=\frac{1}{3}-\frac{a\Delta}{2}\ge 0$, $w_{2a+2}=w_{2a+4}=w_{2a+6}=0$, and $w_i=0$ for all $2a+7\le i\le n$.
  By assumption we have $n\ge 2a+6$ and the first $2a$ voters are obviously passers. By checking $0\le \frac{1}{3}-\frac{a\Delta}{2}<\frac{\Delta}{2}$ 
  we conclude that the remaining voters are null voters and have a non-negative weight. By construction, the weights of the $n$ voters 
  sum up to one. Changing the weights of player $2i-1$ and player $2i$ for $1\le i\le a$ does not change the game so that we obtain 
  a second representation with quota $q$ and weights $\overline{w}_{2i}=\Delta$, $\overline{w}_{2i-1}=\Delta/2$ for $1\le i\le a$, 
  $\overline{w}_{2a+2}=\overline{w}_{2a+4}=\overline{w}_{2a+6}=\frac{1}{3}-\frac{a\Delta}{2}\ge 0$, 
  $w_{2a+1}=w_{2a+3}=w_{2a+4}=\overline{w}_{2a+1}=\overline{w}_{2a+2}=\overline{w}_{2a+3}=0$, and $\overline{w}_i=0$ for all $2a+7\le i\le n$. 
  With this, we have $\Vert w-\overline{w}\Vert_{1}=a\Delta +2-3a\Delta=2(1-a\Delta)\ge\frac{2}{3}$ and $\Vert w-\overline{w}\Vert_{\infty}=\Delta/2$. 
\end{Proof}   
For each $w,\overline{w}\in\mathbb{R}^n$ with $\Delta(w)=\Delta(\overline{w})$, we obviously have $\Vert w-\overline{w}\Vert_\infty\le \Delta(w)$. 
So, a constant lower bound for the $\vert\cdot\Vert_\infty$-norm can only exist if we slightly weaken the assumptions. 

\begin{Lemma}
  \label{lemma_lb_approximation_delta_2}
  For $\Delta\in(0,1]$ there exists a weighted game $v=[q;w]=[\overline{q};\overline{w}]$ with $n\ge \frac{1}{\Delta}+1\ge 2$, 
  $q,\overline{q}\in(0,1)$, $w,\overline{w}\in\mathbb{R}^n_{\ge 0}$, $\Delta(w)=\Delta$, and  
  $\Vert w\Vert_1=\Vert \overline{w}\Vert_1=1$ such that $\Vert w-\overline{w}\Vert_{\infty}\ge\frac{1}{3}$.
\end{Lemma}
\begin{Proof}
  If $1\ge \Delta\ge \frac{2}{3}$ we can apply Lemma~\ref{lemma_lb_approximation_delta_1} or the subsequent example, so that we assume 
  $\Delta<\frac{2}{3}$ in the following. With $a:=\left\lfloor 1/\Delta\right\rfloor$ we set $w_i=\Delta$ for $1\le i\le a$, 
  $w_{a+1}=1-a\Delta$, and $w_i=0$ for $a+2\le i\le n$. Note that $a\ge 1$, $0\le w_{a+1}<\Delta$, $\Delta(w)=\Delta$, $\Vert w\Vert_1=1$, and 
  $a+1\le \frac{1}{\Delta}+1$. 
  
  If $w_{a+1}>0$ we set $q=w_{a+1}$, so that all voters $1\le i\le a+1$ are passers and the remaining voters are null voters.  
  Another representation of the same is given by $\overline{q}=\varepsilon/a$, $\overline{w}_1=1-\varepsilon$, 
  $\overline{w}_i=\varepsilon/a$ for $2\le i\le a+1$, and $\overline{w}_i=0$ for $a+2\le i\le n$, where $\varepsilon=2/3-\Delta>0$. 
  By construction all weights $\overline{w}_i$ are non-negative, $\Vert \overline{w}\Vert_1=1$, and 
  $\Vert w-\overline{w}\Vert_{\infty}\ge 1-\Delta-\varepsilon\ge \frac{1}{3}$.  
  
  If $w_{a+1}=0$ we set $q=\Delta$, so that all voters $1\le i\le a$ are passers and the remaining voters are null voters. 
  Note that $w_{a+1}=1-a\Delta=0$ implies $a\ge 2$. Another representation of the same game is given by $\overline{q}=\varepsilon/(a-1)$.
  $\overline{w}_1=1-\varepsilon$, $\overline{w}_i=\varepsilon/(a-1)$ for $2\le i\le a$, and $\overline{w}_i=0$ for 
  $a+1\le i\le n$, where $\varepsilon=2/3-\Delta>0$. By construction all weights $\overline{w}_i$ are non-negative, 
  $\Vert \overline{w}\Vert_1=1$, and $\Vert w-\overline{w}\Vert_{\infty}\ge 1-\Delta-\varepsilon\ge \frac{1}{3}$. 
\end{Proof}

Combining Lemma~\ref{lemma_general_lower_distance_bound} with lemmas \ref{lemma_lb_approximation_q} and \ref{lemma_lb_approximation_delta_1} gives:
\begin{Proposition}
  \label{prop_impossible}
  Let $\varphi$ be a power index, i.e., a mapping from the set of weighted games (on $n$ voters) into $\mathbb{R}_{\ge 0}^n$.
  \begin{itemize}
    \item[(i)] For each $q\in(0,1]$ and each integer $n\ge 2$ there exists a weighted game $v$ with $n$ voters that 
               permits a representation $[q;w]=v$, where $w\in\mathbb{R}^n_{\ge 0}$ and $\Vert w\Vert_1=1$, 
               such that $\Vert w-\varphi([q;w])\Vert_{1}\ge \frac{1}{3}$ and $\Vert w-\varphi([q;w])\Vert_{\infty}\ge \frac{1}{6}$.
    \item[(ii)] For each $\Delta\in(0,1)$ and each integer $n\ge \frac{4}{3\Delta}+6$ there exists a weighted game $v$ with $n$ voters that 
               permits a representation $[q;w]=v$, where $q\in(0,1]$, $w\in\mathbb{R}^n_{\ge 0}$, $\Vert w\Vert_1=1$, and $\Delta(w)=\Delta$, 
               such that $\Vert w-\varphi([q;w])\Vert_{1}\ge \frac{1}{3}$, and $\Vert w-\varphi([q;w])\Vert_{\infty}\ge\Delta/4$.
  \end{itemize} 
\end{Proposition}


So, we have shown that controlling either the relative quota or the maximum relative weight is not sufficient to obtain reasonable 
upper bounds for the distance between relative weights and power if the number of voters gets large. 
However, it is sufficient to control the quota $q$ and the maximum relative weight $\Delta$ for some power indices as we will see in the 
next subsection. (If $\Delta$ tends to zero, then 
the number of voters automatically tends to $\infty$ since $\Delta\ge\frac{1}{n}$. Due to Lemma~\ref{lemma_relation_maximum_laakso_taagepera} 
it would also be sufficient to control the quota and the Laakso-Taagepera index.) 

In some applications only weighted games with a quota of at least one half are considered which clashes with some of our constructions 
in the proofs of the previous lemmas. However, by considering the dual of a given weighted game we can turn a quota below one half to 
a quota above one half. So, instead of small quotas we get large quotas. 

\subsection{Upper bounds for the distance between weights and power}
\label{subsec_3_3}

We start with a rather general upper bound for all positive and efficient power indices $\varphi$ satisfying 
$\sum_{i\in S} \varphi_i([q;w)]\ge q$ for every winning coalition $S$. This directly implies an upper bound for the nucleolus.  

\begin{Lemma}
  \label{lemma_qdelta_bound_winning}
  Let $w\in\mathbb{R}^n_{\ge 0}$ with $\Vert w\Vert_1=1$ for an integer $n\in\mathbb{N}_{>0}$ and $0<q<1$. For each $x\in\mathbb{R}^n_{\ge 0}$ with 
  $\Vert x\Vert_1=1$ and $x(S)=\sum_{s\in S}x_s\ge q$ for every winning coalition $S$ of $[q;w]$, we have
  $
    \Vert w-x\Vert_1 \le \frac{2\Delta}{\min\{q+\Delta,1-q\}}\le\frac{2\Delta}{\min\{q,1-q\}}
  $, 
  where $\Delta=\Delta(w)$.
\end{Lemma}
\begin{Proof}
  Consider a winning coalition $T$ such that $x(T)$ is minimal and invoke $x(T)\ge q$, see Section~\ref{sec_evacuated_proofs} 
  for the technical details. 
\end{Proof}

\begin{Corollary}
  \label{cor_nucleus}
  Let $w\in\mathbb{R}^n_{\ge 0}$ with $\Vert w\Vert_1=1$ for an integer $n\in\mathbb{N}_{>0}$ and $0<q<1$. For each 
  element $x$ of the nucleus\footnote{The nucleus of a weighted game $[q;w]$ is the set of all $x\in\mathbb{R}_{\ge 0}^n$ with 
  $\Vert x\Vert_1=1$ that minimize the maximum excess $E_1(x)= \max_{S\subseteq N} v(S)-x(S)$. If $[q;w]$ contains passers, 
  then those $x$ may not be individually rational, i.e., $x_i\ge v(\{i\})$ is violated. This case is excluded by some authors.}, 
  which contains the nucleolus, of $[q;w]$, we have
  $
    \Vert w-x\Vert_1 \le \frac{2\Delta}{\min\{q+\Delta,1-q\}}\le\frac{2\Delta}{\min\{q,1-q\}}
  $, 
  where $\Delta=\Delta(w)$.  
\end{Corollary}
\begin{Proof}
  We have $1-x(S)\le E_1(x)$ for every winning coalition $S$ of $[q;w]$, where $E_1(x)$ is the maximum excess. Since $1-w(S)\le 1-q$ 
  for every winning coalition $S$ of $[q;w]$, we have $1-x(S)\le 1-q$ as the maximum excess is minimized for all elements of the nucleus.  
\end{Proof}

Lemma~\ref{lemma_qdelta_bound_winning} and Corollary~\ref{cor_nucleus} generalize  
\cite[Lemma 1]{kurz2014nucleolus} by weakening the assumptions. From Lemma~\ref{lemma_improved_relation_infty_1} we can directly conclude 
similar bounds for the $\Vert\cdot\Vert_\infty$-distance. 

Some power indices $\varphi$ have the property that $\varphi([q;w])$ is a feasible weight vector 
for a suitable quota $q'$, i.e., $[q;w]=[q';\varphi([q;w])]$. Examples are the \emph{minimum sum representation index}, 
see ~\cite{freixas2014minimum}, or power indices based on averaged representations from \cite{kaniovski2015average,
kaniovski2015representation}. For the Penrose-Banzhaf index, the subclass of spherically separable simple games has this 
property, see \cite{houy2014geometry}. Thus, it is appealing to study upper bounds for the $\Vert\cdot\Vert_1$-distance between 
two relative weight vectors of the same weighted game since these imply upper bounds for the distance between power and 
weights given that the power vector can be completed to a representation. From Lemma~\ref{lemma_qdelta_bound_winning} we can directly 
conclude the following two implications:  

\begin{Corollary}
  Let $w,\overline{w}\in\mathbb{R}^n_{\ge 0}$ with $\Vert w\Vert_1=\Vert \overline{w}\Vert_1=1$ for an integer $n\in\mathbb{N}_{>0}$ and 
  $0<q\le \overline{q}<1$. If $[q;w]=[\overline{q};\overline{w}]$ and $\Delta=\Delta(w)$, then we have 
  $
    \Vert w-\overline{w}\Vert_1 \le\frac{2\Delta}{\min\{q,1-q\}}
  $.
\end{Corollary}

\begin{Corollary}
  Let $w,\overline{w}\in\mathbb{R}^n_{\ge 0}$ with $\Vert w\Vert_1=\Vert \overline{w}\Vert_1=1$ for an integer $n\in\mathbb{N}_{>0}$ and 
  $0<q, \overline{q}<1$. If $[q;w]=[\overline{q};\overline{w}]$, then we have 
  $$
    \Vert w-\overline{w}\Vert_1 \le \max\left\{ \frac{2\Delta(w)}{\min\{q,1-q\}},
    \frac{2\Delta(\overline{w})}{\min\{\overline{q},1-\overline{q}\}}\right\}
    \le\frac{2\Delta(w)}{\min\{q,1-q\}}+\frac{2\Delta(\overline{w})}{\min\{\overline{q},1-\overline{q}\}}.
  $$
\end{Corollary}

Unfortunately, those corollaries do not allow us to derive a bound on $\Vert w-\overline{w}\Vert_1$ which only depends on $q$ and $\Delta(w)$. 
However, we can obtain the following analog of Lemma~\ref{lemma_qdelta_bound_winning} for losing 
instead of winning coalitions.

\begin{Lemma}
  \label{lemma_qdelta_bound_losing}
  Let $w\in\mathbb{R}^n_{\ge 0}$ with $\Vert w\Vert_1=1$, $\Delta=\Delta(w)$, and $0<q<1$. For each $x\in\mathbb{R}^n_{\ge 0}$ with 
  $\Vert x\Vert_1=1$ and $x(S)=\sum_{s\in S}x_s\le q$ for every losing coalition $S$ of $[q;w]$, we have 
  $\Vert w-x\Vert_1 \le\frac{4\Delta}{\min\{q,1-q\}}$. Moreover, if $q>\Delta(w)$, then  
  $
    \Vert w-x\Vert_1 \le \frac{2\Delta}{\min\{q-\Delta,1-q+\Delta\}}\le\frac{2\Delta}{\min\{q-\Delta,1-q\}}
  $.
\end{Lemma}
\begin{Proof}
  Consider a losing coalition $T$ such that $x(T)$ is maximal and invoke $x(T)\le q$. Section~\ref{sec_evacuated_proofs} 
  provides technical details. 
\end{Proof}

Intuitively, the inconspicuous condition $\Delta(w)<q$ is equivalent to the property that $[q;w]$ does not contain passers.
\begin{Corollary}
  \label{cor_weighted_representation}
  Let $w,\overline{w}\in\mathbb{R}^n_{\ge 0}$ with $\Vert w\Vert_1=\Vert \overline{w}\Vert_1=1$, $\Delta=\Delta(w)$, and 
  $0<q, \overline{q}<1$. If $[q;w]=[\overline{q};\overline{w}]$, then we have $\Vert w-\overline{w}\Vert_1\le \frac{4\Delta}{\min\{q,1-q\}}$. 
  Moreover, if additionally $[q;w]$ does not contain any passer, then we have $\Vert w-\overline{w}\Vert_1\le$ 
  $$
     \min \left\{\frac{2\Delta}{\min\{q-\Delta,1-q\}},
    \frac{2\Delta(\overline{w})}{\min\{\overline{q}-\Delta(\overline{w}),1-\overline{q}\}}\right\} 
    \le \frac{2\Delta}{\min\{q-\Delta,1-q\}}.
  $$
\end{Corollary}
\begin{Proof}
   If $\overline{q}\ge q$, then $\overline{w}(S)\ge \overline{q}\ge q$ for every winning coalition $S$ of $[q;w]$. Here, we can apply 
   Lemma~\ref{lemma_qdelta_bound_winning}. Otherwise we have $\overline{w}(T)<\overline{q}<q$ for every losing coalition $T$ of $[q;w]$ 
   and Lemma~\ref{lemma_qdelta_bound_losing} applies. For the second bound note that we can interchange the roles of $(q,w)$ and 
   $(\overline{q},\overline{w})$ and take the tighter of the two resulting bounds.
\end{Proof}

\medskip

From Corollary~\ref{cor_weighted_representation} we can deduce the following:
\begin{Theorem}
  \label{thm_main}
  Let $w\in\mathbb{R}^n_{\ge 0}$ with $\Vert w\Vert_1=1$ and $0<q<1$. 
  If a power index $\varphi$ permits the existence of a quota $q'\in(0,1)$ such that $[q';\varphi([q;w])]=[q;w]$, i.e., 
  that the power vector of the given weighted game can be completed to a representation of the same game, then
  $$
    \Vert w-\varphi([q;w])\Vert_1
    \le \frac{4\Delta(w)}{\min\{q,1-q\}}.
  $$   
\end{Theorem}
As mentioned before, \textit{representation compatibility} of $\varphi$ for $[q;w]$ is automatically satisfied
for the minimum sum representation index or one of the power indices based on averaged representations for all weighted games 
and for the Penrose-Banzhaf index for all spherically separable simple games. The theorem also applies to the bargaining model for
weighted games analyzed in \cite{market_value_model}, cf.~\cite{prop_payoffs}.

We remark that Lemma~\ref{lemma_qdelta_bound_winning} and Lemma~\ref{lemma_qdelta_bound_losing} are also valid for roughly weighted games, 
where coalitions with a weight sum being equal to $q$ may also be losing. So, one might ask the same question for $\alpha$-roughly weighted games, 
see \cite{freixas2014alpha,gvozdeva2013three}, where coalitions with weight sum below $q$ are losing and coalitions with weight sum  
above $\alpha q$ are winning. 

\medskip

In the previous subsection we have argued that reasonable upper bounds on the distance between weights and power are impossible 
if only the relative quota or the maximum relative weight is taken into account. If both invariants are known, we have presented 
corresponding upper bounds for some power indices. So far we know that both invariants have to be involved in every upper bound 
somehow, but the tightest possible functional correlation is unknown. To that end, we provide the following lower bound matching the 
shape of the upper bound. 

\begin{Lemma}
  \label{lemma_lb_diam_representation_polytop}
  For each $0<\hat{q}<1$, $0<\Delta<1$ and each sufficiently large integer $n$ there exist weight vectors  
  $w,\overline{w}\in\mathbb{R}^n_{\ge 0}$ with $\Vert w\Vert_1=\Vert \overline{w}\Vert_1=1$, $\Delta(w)=\Delta$ and a quota 
  $0<\overline{q}<1$ with $[\hat{q};w]=[\overline{q};\overline{w}]$ such that 
  $\Vert w-\overline{w} \Vert_1\ge c\cdot \min\left\{2, \frac{\Delta}{\min\{\hat{q},1-\hat{q}\}}\right\}$, where $\Delta=\Delta(w)$ 
  and $c=\frac{1}{5}$. 
\end{Lemma}
\begin{Proof}
  A construction of a matching representation $(\overline{q},\overline{w})$ is provided in Section~\ref{sec_evacuated_proofs}. 
\end{Proof}

Via Lemma~\ref{lemma_general_lower_distance_bound} this can be turned into:
\begin{Theorem}
  \label{thm_lower_bound}
  Let $\varphi$ be a power index, i.e., a mapping from the set of weighted games (on $n$ voters) into $\mathbb{R}_{\ge 0}^n$. 
  For each weighted game $v=[q;w]$, where $\Vert w\Vert_1=1$, there exists a normalized representation $(\overline{q};\overline{w})$ such that
  $$\Vert \overline{w}-\varphi(v) \Vert_1\ge \frac{1}{10}\cdot \min\left\{2, \frac{\Delta(w)}{\min\{q,1-q\}}\right\}.$$  
\end{Theorem}

So, the upper bounds of 
Lemma~\ref{lemma_qdelta_bound_winning} and Corollary~\ref{cor_weighted_representation} are tight up to the involved constant $c$. 

\begin{table}[htp!]
  \begin{center}
    \begin{tabular}{cccccccc}
       \hline
           & Shapley- & Penrose- &          & Public & Deegan-&       & Shift- \\
       $n$ & Shubik   & Banzhaf  & Johnston & Good   & Packel & Shift & DP     \\
       \hline
       3 & 0.33333 & 0.20000 & 0.50000 & 0.33333 & 0.00000 & 0.33333 & 0.00000 \\
       4 & 0.50000 & 0.40000 & 0.75000 & 0.51429 & 0.30000 & 0.51429 & 0.50000 \\
       5 & 0.60000 & 0.57895 & 0.87500 & 0.70330 & 0.50000 & 0.80769 & 0.75000 \\
       6 & 0.66667 & 0.72222 & 1.00000 & 1.00000 & 0.71795 & 1.25763 & 1.24444 \\
       7 & 0.71429 & 0.82609 & 1.13710 & 1.43590 & 1.16923 & 1.60131 & 1.55556 \\
       8 & 0.75000 & 0.89552 & 1.29167 & 1.78649 & 1.49020 & 2.13108 & 2.08929 \\
       9 & 0.77778 & 0.98154 & 1.49796 & 2.01504 & 1.71429 & 2.53762 & 2.43750 \\ 
       \hline
       \phantom{a}
    \end{tabular}
    \caption{Necessary constant $c$ for the approximation of the normalized minimum sum integer representation.}
    \label{table_numerical_results}
  \end{center}
\end{table}

In order to prove upper bounds similar to Theorem~\ref{thm_main} for power indices that cannot be completed to a representation in general, 
it suffices to consider an arbitrary weighted representation for each weighted game, since 
we can use Corollary~\ref{cor_weighted_representation} and the triangle inequality to transfer the result to any other 
weighted representation (while, of course, the involved constant of the upper bound has to be increased). We can use 
that insight also in the other direction, i.e., to numerically check whether such an upper bound 
for a given power index might exist at all. 
Table~\ref{table_numerical_results} lists the maximum necessary constant $c$ so that 
$\Vert \varphi([q;w])-w\Vert_1\le\frac{c \cdot \max_i w_i}{\min\{q,1-q\}}$ for each weighted game with $n$ voters. 
As representation $(q;w)$ we have chosen the normalization of a minimum sum integer representation, see 
e.g.~\cite{kurz2012minimum}. There are 993\,061\,482 weighted games with $n=9$ voters, see~\cite{kartak2015minimal}. 
The exact numbers are unknown for $n>9$. For the definitions of the considered power indices, i.e., Shapley-Shubik, Penrose-Banzhaf, Johnston, 
Public Good, Deegan-Packel, Shift and Shift-Deegan-Packel index, we refer, e.g., to 
\cite{kurz2016inverse}. Actually, we included all power indices known to us, even the scarcely applied ones, that have no upper 
bound implied by the previous results from this section.

The, admittedly, sparse and possibly biased data of Table~\ref{table_numerical_results} suggests that there may be no result 
like Theorem~\ref{thm_main} for the latter five power indices. Let us have a closer look at the most commonly used power indices, 
i.e., the Shapley-Shubik and the Penrose-Banzhaf index. For the Shapley-Shubik index the {\lq\lq}worst case{\rq\rq} examples in the 
setting of Table~\ref{table_numerical_results} can be easily guessed.

\begin{Lemma}
  \label{lemma_worse_case_ssi}
  For $n\ge 3$, $v=[n-1;n-1,\overset{n-1}{\overbrace{1,\dots,1}}]$, $\overline{q}=\frac{1}{2}$, 
  $\overline{w}=\left(\frac{1}{2},\frac{1}{2n-2},\dots,\frac{1}{2n-2}\right)$ 
  where $v=[\overline{q};\overline{w}]$ and $\Vert\overline{w}\Vert_1=1$, we have 
  $\varphi(v)=\frac{1}{n(n-1)}\cdot \left((n-1)^2,1,\dots,1\right)$ and 
  $\Vert \varphi(v)-\overline{w}\Vert_1=\frac{n-2}{n}$ for the Shapley-Shubik index $\varphi$. 
\end{Lemma}
\begin{Proof}
  For a voter $2\le i\le n$ we only need to consider the winning coalition $S=\{2,\dots,n\}$, so that  
  $\varphi_i(v)=\frac{(n-2)!\cdot 1!}{n!}=\frac{1}{n(n-1)}$ and $\varphi_1(v)=1-\sum_{i=2}^n \varphi_i(v)=\frac{n-1}{n}$.
\end{Proof}
We conjecture that Lemma~\ref{lemma_worse_case_ssi} gives indeed the worst case scenario for the Shapley-Shubik index. If true, this 
would imply $\Vert w-\varphi([q;w])\Vert_1 \le \frac{5\Delta(w)}{\min\{q,1-q\}}$ for the Shapley-Shubik index and weighted games 
$[q;w]$ in normalized representation.

\begin{figure} [htp]
  \begin{center}
    \includegraphics[width=10cm]{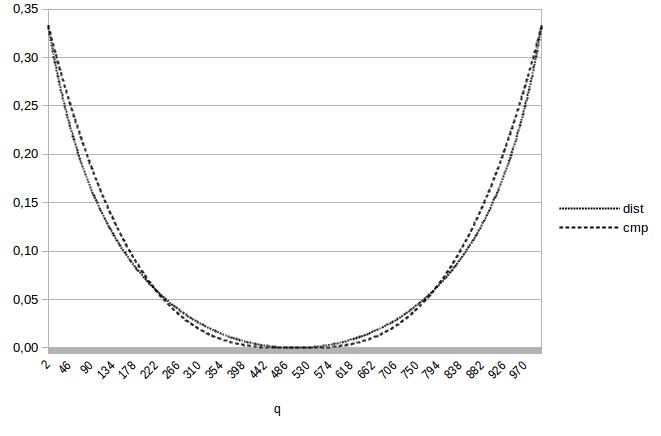}
    \caption{Deviation between weights and the Penrose-Banzhaf index.}
    \label{fig_bzi}
  \end{center}
\end{figure}

For the 
Penrose-Banzhaf index the very same example leads to the power distribution $\frac{1}{2^{n-1}+n-2}\cdot \left(2^{n-1}-1,1,\dots,1\right)$, 
so that the corresponding constant $c$ quickly tends to $1$. While this indeed gives the worst case example for $n\le 8$, things get worse 
for larger $n$.  
To that end, let $w_i=2$ for $1\le i\le m$, $w_i=1$ for $m+1\le i\le 2m$, and $q=\alpha\cdot 3m$, 
where $m\ge 1$ and $\alpha\in (0,1)$. If $\overline{q}(m)$ and $\overline{w}(n)$ denote the normalized quota and weights, 
then the limit $\lim_{m\to\infty} \Vert \varphi([\overline{q}(m);\overline{w}(m)]-\overline{w}(m)\Vert_1$ exists for the 
Penrose-Banzhaf index $\varphi$. Note that $\overline{q}(m)=\alpha$. We have depicted the corresponding limits for different 
values of $\overline{q}(m)=\alpha$ in Figure~\ref{fig_bzi} as \texttt{dist}, where the $x$-axis is labeled with $1000\overline{q}=1000\alpha$ and 
the $y$-axis with the $\Vert\cdot\Vert_1$-distance. We remark that the function is symmetric to $\alpha=\frac{1}{2}$ and takes values 
between zero and $\tfrac{1}{3}$. As a close approximation we have plotted the function $f(x)=\left|x-\frac{1}{2}\right|^3\cdot\frac{8}{3}$ 
labeled as \texttt{cmp}. So, the $\Vert\cdot\Vert_1$-distance between relative weights and the corresponding power distribution according 
to the Penrose-Banzhaf index converges to a constant while the maximum relative weight $\Delta$ tends to zero for a fixed 
relative quota. There are only two types of voters with shares of $\tfrac{2}{3}$ and $\tfrac{1}{3}$, respectively. This example shows that 
it is not possible to derive a general PLT-type result for the Penrose-Banzhaf index if the relative quota does not equal $\tfrac{1}{2}$. 
In that direction numerical simulations and analytical results can be found in \cite{chang2006ls} and \cite{lindner2007cases}, respectively.  
For the other power indices from Table~\ref{table_numerical_results}, besides the Shapley-Shubik index, similar deviations occur for the same 
example.     

For $q=\frac{1}{2}$ we are not aware of examples that would prevent an upper bound in the form of Inequality~(\ref{ie_general_form}). 
Another way to circumvent the suggested non-existence results is to restrict the class of weighted games to special subclasses like e.g.\ 
spherically separable games for the Penrose-Banzhaf index. Another example is given by the rather narrow class of unanimity or quota games 
which can be parameterized as $[q;1,\dots,1]$ for some integer $1\le q\le n$ for $n$ voters. Any symmetric and efficient power index 
assigns a power of $\frac{1}{n}$ to every voter, so that Theorem~\ref{thm_main} can be applied. 

\section{Implications of upper bounds on the $\Vert\cdot\Vert_1$-distance}
\label{sec_implications}

Several results in the literature about limit results for the power distribution of weighted voting games are not given in the 
form of Inequality~(\ref{ie_general_form}). However, we will argue in this subsection that several formulations can be concluded 
from a given upper bound in the shape of Inequality~(\ref{ie_general_form}).

If the $\Vert\cdot\Vert_1$-distance between power and weights is small and the relative weight sum of all voters 
sharing the same weight as voter $i$ does not vanish, then the quotient between power and relative weight for voter~$i$ has to be 
close to $1$. 

\begin{Lemma}
  \label{lemma_quotient}
  Let $w\in\mathbb{R}_{\ge 0}^n$ with $\Vert w\Vert_1=1$, $0<q<1$ and $\varphi$ be a symmetric, efficient, and positive power index.
  If $\Vert \varphi([q;w])-w\Vert_1\le \varepsilon$, then 
  $$
    1-\frac{\varepsilon}{2\alpha}\le \frac{\varphi_i([q;w])}{w_i} \le 1+\frac{\varepsilon}{2\alpha}
  $$
  for all $1\le i\le n$ with $w_i>0$, where $\alpha\:=w(S)>0$ and $S:=\{ 1\le j\le n\,:\,w_i=w_j\}$.
\end{Lemma}
\begin{Proof}
  Assuming $\frac{\varphi([q;w])_i}{w_i} > 1+\frac{\varepsilon}{2\alpha}$ 
  yields $\varphi(S)-w(S)>\varepsilon/2$ by 
  summing over all $j\in S$. Since $\Vert\varphi\Vert_1=\Vert w\Vert_1=1$ and $\varphi,w\in\mathbb{R}_{\ge 0}^n$ we would have 
  $\Vert w-\varphi\Vert_1>\varepsilon$ -- a contradiction. Assuming $\frac{\varphi([q;w])_i}{w_i} < 1-\frac{\varepsilon}{2\alpha}$ 
  yields $w(S)-\varphi(S)>\varepsilon/2$, which leads to the same contradiction.
\end{Proof}

Using the mass measure $\alpha$ is necessary since for each integer $n\ge 2$ we may consider the weighted game $v$ 
consisting of $n-1$ voters of weight $2$ and one voter of weight $1$. Let $\varphi$ be a symmetric and efficient power index that satisfies 
the null voter property and $\overline{w}$ denote the corresponding relative weights. If the quota $q$ is an odd integer, we have 
$\varphi_i(v)=\frac{1}{n}$ for all $1\le i\le n$, so that $\Vert \varphi(v)-\overline{w}\Vert_1=\frac{n-1}{n}\cdot \frac{2}{2n-1}$. 
If $q$ is an even integer, then the voter with weight $1$ is a null voter and all other voters get $\varphi_i(v)=\frac{1}{n-1}$ due to 
symmetry and efficiency. Here we have $\Vert \varphi(v)-\overline{w}\Vert_1=\frac{2}{2n-1}$. So $\Vert \varphi(v)-\overline{w}\Vert_1$ tends 
to zero as the number $n$ of voters approaches infinity. However, the fraction $\frac{\varphi_n(v)}{\overline{w}_i}$ is either $0$ or 
$2-\frac{1}{n}$, i.e., rather far away from $1$ for larger $n$.  


Bounds for quotients between power and weights for two involved players can be deduced from Lemma~\ref{lemma_quotient} 
via:

\begin{Lemma}
  If $w_i,w_j,\varphi_i,\varphi_j\in\mathbb{R}_{>0}$, $\varepsilon_i,\varepsilon_j\in[0,1)$ with 
  $1-\varepsilon_i\le\frac{\varphi_i}{w_i}\le 1+\varepsilon _i$ 
  and $1-\varepsilon_j\le\frac{\varphi_j}{w_j}\le 1+\varepsilon_j$, then
  $$
    \frac{1-\varepsilon_i}{1+\varepsilon_j}\le \frac{w_i}{w_j}\cdot \frac{\varphi_j}{\varphi_i}\le \frac{1+\varepsilon_i}{1-\varepsilon_j}
    \quad\text{and}\quad
    \left|\frac{\varphi_i}{w_i}-\frac{\varphi_j}{w_j}\right|\le \varepsilon_i+\varepsilon_j. 
  $$
\end{Lemma}

\section{Conclusions and directions for further research}
\label{sec_conclusion}

If one is interested in upper bounds on the distance between relative weights and a corresponding power distribution or limit 
results for sequences of weighted games, given the relative quota and the maximum relative weight, Inequality~(\ref{ie_general_form}) 
is the essential answer. In Section~\ref{sec_implications} we have shown that related results can be generally concluded. 

We have derived upper bounds in the form of Inequality~(\ref{ie_general_form}) for the nucleolus and all power indices 
based on weighted representations. Additionally, we have shown that for an arbitrary power index it is not possible to obtain 
bounds of smaller magnitude. So, future technical contributions might try to decrease the corresponding constants $c$ as far as possible. 
This contribution traded smaller constants for easier proofs. An important open problem is 
whether the Shapley-Shubik index permits an upper bound of the form of Inequality~(\ref{ie_general_form}) or if the relation 
between $q$ and $\Delta$ has to be of a different shape. While we showed that the monotonicity behavior of the 
Laakso-Taagepera index is the same as for the inverse maximum relative weight, it might still be possible that the 
Laakso-Taagepera index permits tighter bounds than the maximum relative weight. Moreover, it seems worthwhile to study 
other invariants than those used here.

Regarding limit results we showed that the power distribution tends to the relative weights under the $\Vert\cdot\Vert_1$-distance 
for the nucleolus and power indices based on weighted representations as long as the maximum relative weight tends 
to zero and the quotas are not too skewed (i.e., bounded away from $0$ and $1$.) An analytical example with 
two types of voters having non-vanishing weight shares illustrates that the $\Vert\cdot\Vert_1$-distance between the 
Penrose-Banzhaf power distribution and the relative weights tends to a strictly positive number provided that the quota 
is a fixed number different from $\frac{1}{2}$. So, for the Penrose-Banzhaf index power can not converge to weights provided the relative 
quota is not pegged at one half. Besides the Shapley-Shubik index, for which the general convergence was proven in 
\cite{neyman1982renewal}, there seems to be no suitable candidate for another such limit result among the classical power indices 
if the class of weighted games is not restricted to suitable subclasses or only voters of specific types are considered in PLT type results. 
Weakening the assumptions may allow more positive results.

For a weighted voting game $[q;w]$ with many voters and a power index $\varphi$ that permits an upper bound like Inequality~(\ref{ie_general_form}) 
the exact evaluation of $\varphi([q;w])$ may be replaced by the computation of a normalized representation at the cost of an 
approximation error that can be upper bounded by a concrete formula. Since Inequality~(\ref{ie_general_form}) does not involve 
the weights directly we may also apply this approximation in the case where some weights are unknown. However, there is one 
major drawback. The presented results give reasonable upper bounds for the approximation error only in those cases where the maximum 
relative weight is comparatively small. 

The case of the maximum relative weight going to zero one sometimes speaks of an \textit{ocean} of voters.         
In contrast, some voters, called \textit{islands}, have a relatively large weight and all others have an 
individual weight that is comparatively negligible. However, this is not the case for the weight sum of all {\lq\lq}small{\rq\rq} 
voters. Limit results are available for the Shapley-Shubik and the Penrose-Banzhaf index in the oceanic world with a finite number of atoms, see 
\cite{dubey1979mathematical}. Our example on the ownership structure of a stock corporation from the introduction 
may very likely belong to the oceanic world with a few islands if large stockholders like e.g.\ hedge funds are involved. \cite{leech2013power} 
reported that limit results for both cases are commonly stated separately and that a unification is needed. 
Lemma~\ref{lemma_relation_maximum_laakso_taagepera} and Corollary~\ref{cor_relation_maximum_laakso_taagepera} 
show very transparently that the distinction using the Laakso-Taagepera index is the very same as the distinction using the maximum 
relative weight. Lemma~\ref{lemma_lb_diam_representation_polytop} rather shows that 
it is impossible to derive meaningful bounds in the oceanic case with islands given the premises underlying Inequality~(\ref{ie_general_form}).  
However, it seems reasonable to assume that the weights of the \textit{large} voters are known 
with high accuracy and that their number is relatively small. This would allow to make use of combinatorial algorithms. The idea 
is to solve an auxiliary problem to compute an approximation for the power distribution of the large voters. Suppose that for a 
set $N$ of voters we classify the voters into many small ones, collected in $O$, and a few large ones, collected in $N\backslash O$. 
Let $\overline{w}$ be the vector of relative weights, $\overline{q}$ be the relative quota, $\alpha=\overline{w}(O)$ 
the weight mass of the small voters, and $x$ be an optimal solution of
\begin{eqnarray*}
  \min y &\!\!\!& \text{subject to
}\\
  y+\sum_{i\in S}x_i\ge 1 && \forall S\subseteq N\backslash O:\, \overline{w}(S)\ge \overline{q}\\
  y+ \frac{\overline{q}-\overline{w}(S)}{\alpha}\cdot \beta+  \sum_{i\in S}x_i\ge 1 && \forall S\subseteq N\backslash O:\, \overline{q}-\alpha\le\overline{w}(S)<\overline{q}\\
  \beta+\sum_{i\in N\backslash O} x_i =1\\
  x_i\in\mathbb{R}_{\ge 0}&&\forall i\in N\backslash O\\
  \beta\in\mathbb{R}_{\ge 0} 
\end{eqnarray*}
We claim that $x_i^\star=x_i$ for $i\in N\backslash O$ and $x_i^\star=\overline{w_i}\cdot\frac{\beta}{\alpha}$ is a good approximation 
for the nucleolus $\overline{x}$ of $[\overline{q};\overline{w}]$. More precisely, we conjecture that there exists a constant 
$c\in\mathbb{R}_{>0}$ such that
$$
  \Vert \overline{x}-x^\star\Vert_1\le \frac{c\Delta_O}{\min\left \{\left |\overline{q}-\overline{w}(S)\right|\,:\,S\subseteq N\backslash O\right\}},
$$ 
where $\Delta_O=\max\{\overline{w}_i\,:\, i\in O\}$ is the maximum relative weight of a small voter. The idea is to treat the small 
voters as a continuum and to determine a vector that minimizes the maximum excess. This is the first step of the optimization 
problem for the nucleolus. Preliminary  results in the direction of this conjecture were obtained in \cite{galil1974nucleolus} quite some years 
ago. For suitable auxiliary problems for the Shapley-Shubik and the Penrose-Banzhaf index we refer to \cite{dubey1979mathematical}. 
The proposed direction of research of this last paragraph departs from the topic of this paper, i.e., the approximation of power by weights. 
However, it suggests to look for alternative approximations that can be applied once the number of voters is large or partly unknown 
or uncertain in the oceanic case with islands where the obtained results on the deviation between power and weights are impractical.

\section*{Acknowledgment}
\noindent
The author wishes to thank Alexander Mayer and three anonymous reviewers for his or her comments on an earlier draft of this paper.

\appendix

\section{Delayed proofs}
\label{sec_evacuated_proofs}
\begin{Proof} \textbf{(Lemma~\ref{lemma_relation_maximum_laakso_taagepera})}\\
  For $n=1$, we have $w_1=1$, $\Delta(w)=1$, $\alpha=0$, and $L(w)=1$, so that we assume $n\ge 2$ in the remaining part of the proof.
  For $w_i\ge w_j$ consider $a:=\frac{w_i+w_j}{2}$ and $x:=w_i-a$, so that $w_i=a+x$ and $w_j=a-x$. With this we have 
  $w_i^2+w_j^2=2a^2+2x^2$ and $(w_i+y)^2+(w_j-y)^2=2a^2+2(x+y)^2$. Let us assume that $w^\star$ minimizes $\sum_{i=1}^n w_i^2$ under 
  the conditions $w\in\mathbb{R}_{\ge 0}$, $\Vert w\Vert_1=1$, and 
  $\Delta(w)=\Delta$. (Since the target function is continuous and the feasible set is compact and non-empty, a global minimum indeed 
  exists.) W.l.o.g.\ we assume $w_1^\star=\Delta$. If there are indices $2\le i,j\le n$ with $w_i^\star>w_j^\star$, i.e., $x>0$ in the 
  above parameterization, then we may choose $y=-x$. Setting $w_i':=w_i^\star+y=a=\frac{w_i^\star+w_j^\star}{2}$, $w_j':=w_j^\star-y=a
  =\frac{w_i^\star+w_j^\star}{2}$, and $w_h':=w_h^\star$ for all $1\le h\le n$ with $h\notin\{i,j\}$, we have $w'\in \mathbb{R}_{\ge 0}^n$, 
  $\Vert w'\Vert_1=1$, $\Delta(w')=\Delta$, and $\sum_{h=1}^n \left(w_h'\right)^2=\sum_{h=1}^n \left(w_h^\star\right)^2\,-\,x^2$. 
  Since this contradicts the minimality of $w^\star$, we have $w_i^\star=w_j^\star$ for all $2\le i,j\le n$, so that we conclude 
  $w_i^\star=\frac{1-\Delta}{n-1}$ for all $2\le i\le n$ from $1=\Vert w^\star\Vert_1=\sum\limits_{h=1}^n w_h^\star$.
  Thus, $L(w)\le 1/\left(\Delta^2+\frac{(1-\Delta)^2}{n-1}\right)$, which is tight. Since $\Delta\le 1$ and $n\ge 2$, we have 
  $1/\left(\Delta^2+\frac{(1-\Delta)^2}{n-1}\right)\le \frac{1}{\Delta^2}$, which is tight if and only if $\Delta=1$, i.e., 
  $n-1$ of the weights have to be equal to zero.
  
  Now, let us assume that $w$ maximizes $\sum_{i=1}^n w_i^2$ under the conditions $w\in\mathbb{R}_{\ge 0}$, $\Vert w\Vert_1=1$, and 
  $\Delta(w)=\Delta$. (Due to the same reason a global maximum indeed exists.) Due to $1=\Vert w\Vert_1\le n\Delta$ we have $0<\Delta\le 1/n$, 
  where $\Delta=1/n$ implies $w_i=\Delta$ for all $1\le i\le n$. In that case we have $L(w)=n$ and $\alpha=0$, so that the stated lower bounds 
  for $L(w)$ are valid. In the remaining cases we assume $\Delta>1/n$. If there would exist two indices $1\le i,j\le n$ with $w_i\ge w_j$, 
  $w_i<\Delta$, and $w_j>0$, we may strictly increase the target function by moving weight from $w_j$ to $w_i$ (this corresponds to 
  choosing $y>0$), by an amount small enough to still satisfy the constraints $w_i\le \Delta$ and $w_j\ge 0$. Since $\Delta>0$, we can set 
  $a:=\lfloor 1/\Delta\rfloor\ge 0$ with $a\le n-1$ due to $\Delta>1/n$. Thus, for a maximum solution, we
  have exactly $a$ weights that are equal to $\Delta$, one weight that is equal to $1-a\Delta\ge 0$ (which may indeed 
  be equal to zero), and $n-a-1$ weights that are equal to zero. With this and $a\Delta=1-\alpha\Delta$ we have 
  $
    \sum_{i=1}^n w_i^2=a\Delta^2 (1-a\Delta)^2=\Delta-\alpha\Delta^2+\alpha^2\Delta^2=\Delta(1-\alpha\Delta+\alpha^2\Delta)
    =\Delta\left(1-\alpha(1-\alpha)\Delta\right)\le\Delta
  $.   
  Here, the latter inequality is tight if and only if $\alpha=0$, i.e., $1/\Delta\in\mathbb{N}$.
\end{Proof}

\medskip

\begin{Proof} \textbf{(Lemma~\ref{lemma_qdelta_bound_winning})}\\
  We set $N=\{1,\dots,n\}$, $w(U)=\sum_{u\in U} w_u$ and $x(U)=\sum_{u\in U} x_u$ for each $U\subseteq N$. Let $S^+=\{i\in N \mid x_i>w_i\}$ and 
  $S^-=\{i\in N\mid x_i\le w_i\}$, i.e., $S^+$ and $S^-$ partition the set $N$ of players. We have $w(S^+)<1$ since $w(S^+)<x(S^+)\le x(N)=1$, so 
  that $w(S^-)>0$. Define $0\le \delta\le 1$ by $x(S^-)=(1-\delta)w(S^-)$. We have 
  \begin{equation}
    x(S^+)=1-x(S^-)=w(S^+)+w(S^-)-(1-\delta)w(S^-)=w(S^+)+\delta w(S^-)
  \end{equation}   
  and
  \begin{equation}
    \label{eq_dist_delta}
    \Vert w-x\Vert_1
    =\left(x(S^+)-w(S^+)\right)+\left(w(S^-)-x(S^-)\right)=2\delta w(S^-).
  \end{equation}
  Generate a set $T$ by starting at $T=\emptyset$ and successively add a remaining player $i$ in $N\backslash T$ with minimal
  $x_i/w_i$, where all players $j$ with $w_j=0$ are the worst ones. Stop if $w(T)\ge q$. By construction $T$ is a winning coalition 
  of $[q;w]$ with $w(T)<q+\Delta$, since the generating process did not stop earlier and $w_j\le \Delta(w)$ for all $j\in N$.   
  
  If $w(S^-)\ge q$, we have $T\subseteq S^-$ and $x(T)/w(T)\le x(S^-)/w(S^-)=1-\delta$. Multiplying by $w(T)$ and using $w(T)<q+\Delta$ 
  yields
  \begin{equation}
    x(T)\le (1-\delta)w(T)<(1-\delta)(q+\Delta)= (1-\delta)q +(1-\delta)\Delta. 
  \end{equation}
  Since $x(T)\ge q$, as $T$ is a winning coalition, we conclude $\delta<\Delta/(q+\Delta)$. Using this and $w(S^-)<1$ in Equation~(\ref{eq_dist_delta}) 
  yields 
  \begin{equation}
    \Vert w-x\Vert_1<\frac{2\Delta}{q+\Delta}<\frac{2\Delta}{q}.
  \end{equation}  
  
  If $w(S^-)< q$, we have $S^-\subseteq T$, $x(T)=x(S^-)+x(T\backslash S^-)$, $w(T\backslash S^-)>0$, and $w(S^+)>0$. Since 
  $T\backslash S^-\subseteq S^+$, 
  $x(T\backslash S^-)/w(T\backslash S^-)\le x(S^+)/w(S^+)$, so that
  \begin{eqnarray*}
    x(T)&=&x(S^-)+x(T\backslash S^-)\le (1-\delta)w(S^-)+\frac{x(S^+)}{w(S^+)}\cdot\left(w(T)-w(S^-)\right)\\
    &\le& (1-\delta)w(S^-)+\frac{x(S^+)}{w(S^+)}\cdot\left(q+\Delta-w(S^-)\right)\\
    &=& 
    q+\frac{x(S^+)\Delta-(1-q)\delta w(S^-)}{w(S^+)}\\
    &\le & q+\frac{\Delta-(1-q)\delta w(S^-)}{w(S^+)}.
  \end{eqnarray*} 
  Since $x(T)\ge q$, we conclude $(1-q)\delta w(S^-)\le \Delta$, so that 
  $\Vert w-x\Vert_1\le \frac{2\Delta}{1-q}$.     
\end{Proof}

\medskip

\begin{Proof} \textbf{(Lemma~\ref{lemma_qdelta_bound_losing})}\\
  If $q\le 2\Delta$, then $\frac{4\Delta}{\min\{q,1-q\}}\ge \frac{4\Delta}{q}\ge 2\ge \Vert x-w\Vert_1$, so that we can assume $q>\Delta$. 

  Using the notation from the proof of Lemma~\ref{lemma_qdelta_bound_winning}, we have $x(S^+)=w(S^+)+\delta w(S^-)$ and 
  $\Vert w-x\Vert_1=2\delta w(S^-)$.   
    
  Generate $T$ by starting at $T=\emptyset$ and successively add a remaining player $i$ in $N\backslash T$ with maximal
  $x_i/w_i$, where all players $j$ with $w_j=0$ are taken in the first rounds, as long as $w(T)+w_i< q$. By construction $T$ is a losing coalition 
  of $[q;w]$ with $q-\Delta\le w(T)<q$, since the generating process did not stop earlier. 
  
  If $w(S^+)\ge q$, we have $T\subseteq S^+$ and $x(T)/w(T)\ge x(S^+)/w(S^+)=1+\frac{\delta w(S^-)}{w(S^+)}\ge 1+\delta w(S^-)$. Multiplying 
  by $w(T)$ and using $w(T)\ge q-\Delta$ yields
  \begin{equation*}
    x(T)\ge \left(1+\delta w(S^-)\right)w(T)\ge\left(1+\delta w(S^-)\right)(q-\Delta)= (q-\Delta) +\delta w(S^-)(q-\Delta). 
  \end{equation*}
  Since $x(T)\le q$, as $T$ is a losing coalition, we conclude $\delta w(S^-)\le\Delta/(q-\Delta)$, so that 
  $\Vert w-x\Vert_1<\frac{2\Delta}{q-\Delta}$.
  
  If $w(S^+)< q$, we have $S^+\subseteq T$, $x(T)=x(S^+)+x(T\backslash S^+)$, $w(T\backslash S^+)>0$, and $w(S^-)>0$. Since 
  $T\backslash S^+\subseteq S^-$, 
  $x(T\backslash S^+)/w(T\backslash S^+)\ge x(S^-)/w(S^-)$, so that
  \begin{eqnarray*}
    x(T)&=&x(S^+)+x(T\backslash S^+)\ge  w(S^+)+\delta w(S^-) +\frac{x(S^-)}{w(S^-)}\cdot\left(w(T)-w(S^+)\right)\\
    &\ge& w(S^+)+\delta w(S^-) +(1-\delta)\cdot\left(q-\Delta-w(S^+)\right)\\
    &=& \delta w(S^-)+q-\Delta-\delta q+\delta \Delta+\delta w(S^+)
    = q-\Delta+\delta(1-q+\Delta).
  \end{eqnarray*} 
  Since $x(T)\le q$, 
  $\delta\le \frac{\Delta}{1-q+\Delta}$, so that 
  $\Vert w-x\Vert_1\le \frac{2\Delta}{1-q+\Delta}$ due to $w(S^-)\le 1$.     
  
  So, for $q>\Delta$ we have $\Vert w-x\Vert_1 \le \frac{2\Delta}{\min\{q-\Delta,1-q+\Delta\}}\le \frac{2\Delta}{\min\{q-\Delta,1-q\}}$. 
  In order to show $\Vert w-x\Vert_1\le \frac{4\Delta}{\min\{q,1-q\}}$ it remains to consider the case $q\le 1-q$. For $q>2\Delta$, see 
  the start of the proof, we have $\Vert w-x\Vert_1\le \frac{2\Delta}{\min\{q-\Delta,1-q\}}\le \frac{2\Delta}{q-\Delta}\le 
  \frac{4\Delta}{q}\le \frac{4\Delta}{\min\{q,1-q\}}$. 
\end{Proof}

In order to prove Lemma~\ref{lemma_lb_diam_representation_polytop}, we need two 
preparing lemmas.
 
\begin{Lemma}
  \label{lemma_lb_diam_representation_polytop_part_1}
  For each $0<q\le \Delta\le 1$, each $n\in\mathbb{N}$, and each $0<\varepsilon<\min\{\Delta,1/2\}$ 
  with $n\ge \max\{3,1/\Delta\}$ there exist $w,\overline{w}\in\mathbb{R}_{\ge 0}^n$ 
  with $\Vert w\Vert_1=\Vert\overline{w}\Vert_1=1$, $\Delta(w)=\Delta$, and $0<\overline{q}<1$ with $[q;w]=[\overline{q};\overline{w}]$, 
  such that $\Vert w-\overline{w}\Vert_1\ge 1-2\varepsilon$ and $\Vert w-\overline{w}\Vert_\infty\ge \frac{1}{2}-\varepsilon$.  
\end{Lemma}
\begin{Proof}
  We set $a:=\left\lfloor\frac{1}{\Delta}\right\rfloor\in\mathbb{N}_{\ge 1}$, so that $\frac{1}{2\Delta}<a\le\frac{1}{\Delta}$. 
  
  If $1-a\Delta\ge q$, we set $w_i=\Delta$ for $1\le i\le a$, $w_{a+1}=1-a\Delta$, $w_i=0$ for $a+2\le i\le n$, 
  $\overline{w}_{a+1}=1-\varepsilon$, $\overline{w}_i=\varepsilon/a$ for $1\le i\le a$, $\overline{w}_i=0$ for $a+2\le i\le n$, 
  and $\overline{q}=\varepsilon/a$. Here, we have $\Vert w-\overline{w}\Vert_1=2\cdot \left(a\Delta-\varepsilon\right)>1-2\varepsilon$ 
  and $\Vert w-\overline{w}\Vert_{\infty}=a\Delta-\varepsilon\ge \frac{1}{2}-\varepsilon$. 

  If $1-a\Delta<q$ and $a\ge 2$, then we set $w_i=\Delta$ for $1\le i\le a$, $w_{a+1}=1-a\Delta$, $w_i=0$ for $a+2\le i\le n$, 
  $\overline{w}_1=1-\varepsilon$, $\overline{w}_i=\frac{\varepsilon}{a-1}$ for $2\le i\le a$, $\overline{w}_i=0$ for $a+1\le i\le n$, 
  and $\overline{q}=\frac{\varepsilon}{a-1}$. With this, we have $\Vert w-\overline{w}\Vert_1=1-\varepsilon-\Delta+(a-1)\cdot\left(
  \Delta-\frac{\varepsilon}{a-1}\right)+1-a\Delta=1+(a-2)\Delta-2\varepsilon+(1-a\Delta)\ge 1-2\varepsilon$ and 
  $\Vert w-\overline{w}\Vert_{\infty}\ge 1-\Delta-\varepsilon\ge \frac{1}{2}-\varepsilon$ since $a\ge 2$, so that $\Delta\le\frac{1}{2}$.  

  If $a=1$ and $1-a\Delta=1-\Delta<q$, then we have $\frac{1}{2}<\Delta\le 1$ and we set $w_1=\Delta$, $w_2=1-\Delta$, $w_i=0$ for $3\le i\le n$, 
  $\overline{w}_1=\frac{1}{2}+\varepsilon$, $\overline{w}_2=0$, $\overline{w}_3=\frac{1}{2}-\varepsilon$, $\overline{w}_i=0$ for $4\le i\le n$, 
  and $\overline{q}=\frac{1}{2}$. We have $\Vert w-\overline{w}\Vert_1\ge 1-2\varepsilon$ and $\Vert w-\overline{w}\Vert_\infty\ge 
  \frac{1}{2}-\varepsilon$.
\end{Proof}

\begin{Lemma}
  \label{lemma_lb_diam_representation_polytop_part_2}
  Let $0<\varepsilon<\frac{1}{2}$, $0<q<1$, $b\in\mathbb{N}_{\ge 1}$, $\frac{q}{b+1}\le \Delta<\frac{q}{b}$, and $n\in\mathbb{N}$ 
  with $n\ge \frac{1}{\Delta}+1$. Then, there 
  exist $w,\overline{w}\in\mathbb{R}_{\ge 0}^n$ with $\Vert w\Vert_1=\Vert\overline{w}\Vert_1=1$, $\Delta(w)=\Delta$, and 
  $0<\overline{q}<1$ with $[q;w]=[\overline{q};\overline{w}]$, $\Vert w-\overline{w}\Vert_1> \frac{2}{9}\cdot\frac{\Delta}{q}-\varepsilon$, and 
  $\Vert w-\overline{w}\Vert_{\infty}>\frac{\Delta}{3(\Delta+1)}\cdot \frac{\Delta}{q}-2\Delta\varepsilon$. 
\end{Lemma}
\begin{Proof}
  We set $a:=\left\lfloor\frac{1}{\Delta}\right\rfloor\in\mathbb{N}_{\ge 1}$, so that $\frac{1}{2\Delta}<a\le\frac{1}{\Delta}$. 
  Consider $w_i=\Delta$ for $1\le i\le a$, $0\le w_{a+1}=1-a\Delta<\Delta$, and $w_i=0$ for $a+2\le i\le n$. Observe $\frac{1}{b}>\frac{\Delta}{q}$.
 
  If $b\Delta+1-a\Delta<q$ we set $\kappa:=a$ and $\kappa:=a+1$ otherwise. So, the voters $\kappa+1\le i\le n$ are null voters and the other voters are pairwise equivalent. 
  We have $\kappa\ge 2$ since $\kappa=1$ implies $a=1$ and $\Delta>\frac{1}{2}$, so that $b=1$ and $b\Delta+1-a\Delta=1\ge q$. Additionally, 
  we have $\frac{\Delta}{\Delta+1}\le \frac{1}{\kappa}<2\Delta $. (The right hand side may be decreased to $\frac{3}{2}\Delta$.)
 
  If $\kappa \equiv 0 \pmod 2$, then we set 
  $\overline{w}_i=\frac{2}{\kappa}\cdot\frac{b+1}{2b+1}-\frac{\varepsilon}{\kappa}$ for $1\le i\le \kappa/2$, 
  $\overline{w}_i=\frac{2}{\kappa}\cdot\frac{b}{2b+1}+\frac{\varepsilon}{\kappa}$ for $\kappa/2+1\le i\le \kappa$, 
  $w_i=0$ for $\kappa+1\le i\le n$, and $\overline{q}=\frac{2}{\kappa}\cdot\frac{b^2+b}{2b+1}$. With this, we have $[q;w]=[\overline{q};
  \overline{w}]$, $\Vert \overline{w}\Vert_1=1$, $\Vert w-\overline{w}\Vert_1\ge \frac{1}{2b+1}-\varepsilon> 
  \frac{1}{3}\cdot \frac{\Delta}{q}-\varepsilon$, and 
  $\Vert w-\overline{w}\Vert_{\infty}\ge \frac{1}{\kappa(2b+1)}-\frac{\varepsilon}{\kappa}>\frac{\Delta}{3(\Delta+1)}\cdot \frac{\Delta}{q}-2\Delta\varepsilon$. 
  If instead $\kappa\equiv 1\pmod 2$, then we have $\kappa\ge 3$. In this case we set $\overline{w}_i=\frac{2}{\kappa}\cdot\frac{b+1}{2b+1}-
  \frac{\varepsilon}{\kappa}$ for $1\le i\le (\kappa-1)/2$, $\overline{w}_i=\frac{2}{\kappa}\cdot\frac{b}{2b+1}+\frac{\varepsilon}{\kappa}$ for 
  $(\kappa+1)/2\le i\le \kappa-1$, $\overline{w}_\kappa=\frac{1}{\kappa}$, $w_i=0$ for $\kappa+1\le i\le n$, and 
  $\overline{q}=\frac{2}{\kappa}\cdot\frac{b^2+b}{2b+1}$. With this, we have $[q;w]=[\overline{q};
  \overline{w}]$, $\Vert\overline{w}\Vert_1=1$, $\Vert w-\overline{w}\Vert_1\ge \frac{1}{2b+1}\cdot\left(1-\frac{1}{\kappa}\right)-\varepsilon> 
  \frac{2}{9}\cdot \frac{\Delta}{q}-\varepsilon$, and $\Vert w-\overline{w}\Vert_{\infty}\ge \frac{1}{\kappa(2b+1)}-\frac{\varepsilon}{\kappa}
  >\frac{\Delta}{3(\Delta+1)}\cdot \frac{\Delta}{q}-2\Delta\varepsilon$. 
\end{Proof}

\medskip

\begin{Proof}\textbf{(Lemma~\ref{lemma_lb_diam_representation_polytop})}\\
  If $\hat{q}\le 1-\hat{q}$, we can apply Lemma~\ref{lemma_lb_diam_representation_polytop_part_1} and 
  Lemma~\ref{lemma_lb_diam_representation_polytop_part_2} with $0<\varepsilon<\min\left\{\frac{1}{10},\frac{\Delta}{45\hat{q}}\right\}$. For the 
  other case we remark that the dual of each weighted 
  game $[q;w]$ is given by $[1-q+\tilde{\varepsilon};w]$ for each suitably small $\tilde{\varepsilon}>0$. So, we can   
  apply Lemma~\ref{lemma_lb_diam_representation_polytop_part_1} and Lemma~\ref{lemma_lb_diam_representation_polytop_part_2} 
  with $q=1-\hat{q}+\tilde{\varepsilon}$ and $0<\varepsilon<\frac{1}{2}\cdot \min\left\{\frac{1}{10},\frac{\Delta}{45q}\right\}=:\kappa$. 
  Certainly, we can choose $\tilde{\varepsilon}>0$ small enough to get 
  $$
    \left\Vert \min\left\{2, \frac{\Delta}{\min\{\hat{q},1-\hat{q}\}}\right\}- \min\left\{2, \frac{\Delta}{\min\{q,1-q\}}\right\}\right\Vert_1<\kappa
  $$  
  and $q\le 1-q$. The stated result then follows from the triangle inequality.
\end{Proof}


\end{document}